%% file: main.tex
\documentclass[10pt,journal]{IEEEtran}
\usepackage{cite}
\usepackage{url} 
\usepackage[cmex10]{amsmath}
\usepackage{algorithm}
\usepackage{algorithmic}

\usepackage{array}
\usepackage[caption=false,font=footnotesize]{subfig}
\usepackage{graphicx,listings,psfrag,amsfonts,cases,bm}
\usepackage{pst-sigsys}
\usepackage{pstricks,pst-node,pst-tree}
\usepackage[section]{placeins}
\usepackage{amsbsy}

\input{math_macro.tex}

\newcommand{\subparagraph}{}

\begin{document}
\title{Input Modeling and Uncertainty Quantification for Improving Volatile Residential Load Forecasting}
\author{Guangrui~Xie, 
        Xi~Chen, 
        and~Yang~Weng,~\IEEEmembership{Member,~IEEE}
\thanks{G. Xie and X. Chen are with the Grado Department of Industrial and Systems Engineering, Virginia Tech, Blacksburg,
VA 24061 USA (e-mail: \{guanx92, xchen6\}@vt.edu).}
\thanks{Y. Weng is with the School of Electrical, Computer and Energy Engineering, Arizona State University, Tempe, AZ 85281 USA (e-mail: yang.weng@asu.edu).}
}



\maketitle

\begin{abstract}
Load forecasting has long been recognized as an important building block for all utility operational planning efforts. Over the recent years, it has become ever more challenging to make accurate forecasts due to the proliferation of distributed energy resources, despite the abundance of existing load forecasting methods. In this paper, we identify one drawback suffered by most load forecasting methods --- neglect to thoroughly address the impact of input errors on load forecasts. As a potential solution, we propose to incorporate input modeling and uncertainty quantification to improve load forecasting performance via a two-stage approach. The proposed two-stage approach has the following merits. (1) It provides input modeling and quantifies the impact of input errors, rather than neglecting or mitigating the impact --- a prevalent practice of existing methods. (2) It propagates the impact of input errors into the ultimate point and interval predictions for the target customer's load to improve predictive performance. (3) A variance-based global sensitivity analysis method is further proposed for input-space dimensionality reduction in both stages to enhance the computational efficiency. Numerical experiments show that the proposed two-stage approach outperforms competing load forecasting methods in terms of both point predictive accuracy and coverage ability of the predictive intervals.
\end{abstract}

\section{Introduction}\label{sec:intro}

With the rapid growth of distributed energy resources (DERs), the high level of uncertainty in power grids has been steadily increasing, rendering responsive and reliable load forecasting more critical and challenging than ever. A plethora of methods have been proposed to meet the challenges in load forecasting, which generally fall into two categories: one-stage and two-stage methods. A typical one-stage method adopts a single prediction model for load forecasting, and various models have been considered in the literature. For example, in \cite{Saber2017} and \cite{Amral2007}, standard multiple linear regression models are enhanced by different techniques to tackle large datasets for load forecasting. In \cite{Pappas2008} and \cite{Shao2018}, modified ARIMA models are used to capture customers' load patterns. In \cite{Jiang2018} and \cite{Jain2014Forecast}, support vector regression algorithms are modified to achieve a high predictive accuracy especially when handling high-resolution distribution systems. In \cite{Amjady2006,Rafiei2018,Kong2019}, different types of neural network (NN)-based methods (e.g., recurrent NN, wavelet NN, and fuzzy NN) are proposed for modeling the complex relationships between customers' loads and input variables. Last but not least, Gaussian process (GP)-based methods that adopt various forms of covariance kernels and input variables are studied in \cite{Hachino2014,Alamaniotis2014,Lloyd2014} to obtain an informative interval prediction for the target customer's load.

As the name suggests, two-stage methods, on the other hand, perform load forecasting in two stages, with each stage having a separate model for a different purpose. The second-stage model is dedicated to the target customer's load prediction, while the first-stage model is typically set up to process input features (i.e., either select useful features or provide accurate input estimates) to carry out the second-stage load forecasting. Two-stage methods are known for their enhanced predictive performance thanks to the first-stage processing \cite{Bozic2013}. The interested reader is referred to \cite{Bozic2013,Ghadimi2018,Campion2018,Moon2018,Xie} and references therein for a review of two-stage load forecasting methods.

Despite their successful applications reported, most existing methods have one major drawback: they fail to thoroughly address the impact of input errors when performing point and interval predictions for the target customer's load. Input errors arise from the use of the estimates of relevant input features for load forecasting. The commonly used input features include future weather conditions (e.g., temperature and cloud coverage), future power states (e.g., voltages and phase angles), etc. As these future input values are not known at the time of prediction, they must be estimated, say, by domain experts and prediction models. The predicted input values are inevitably susceptible to errors, which can severely undermine the ultimate load forecasting performance achieved. However, a majority of existing methods neglect or underestimate the corresponding impact as if the input data used for prediction were free of errors.

There indeed exist some two-stage methods that tackle input errors to mitigate their impact on point predictions to some extent. The main idea is to apply noise filtering techniques (e.g., Kalman filter) to filter out input errors. However, this practice does not tackle the impact of input errors on interval predictions, neither does it provide any input uncertainty quantification. A better way is to appropriately quantify the impact of input errors and reflect it in both point and interval load predictions.
 
\begin{figure}[ht]
  \centering%
  \includegraphics[width=8.5cm]{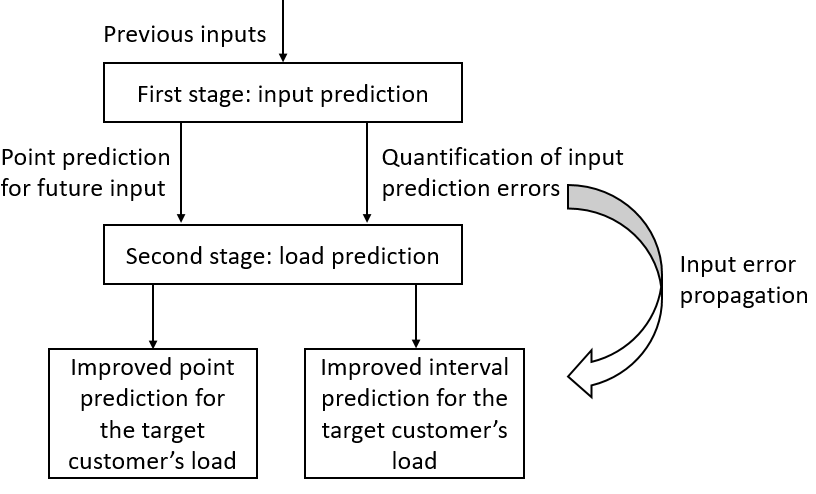}
  \caption{A schematic diagram of the proposed two-stage load forecasting approach.}
  \label{fig:flow}
\end{figure}

Motivated by the aforementioned discussion, we contribute to the load forecasting literature by proposing a two-stage approach to thoroughly address the impact of input errors through input modeling and uncertainty quantification. Specifically, in the first stage, we adopt a hybrid model of NN and GP to provide an accurate point prediction for future input feature values, together with an interval prediction to quantify the input uncertainty. In the second stage, the point and interval predictions obtained are seamlessly utilized by a GP model that can handle noisy inputs, which forms the ultimate point and interval predictions for the target customer's future load with the impact of input errors taken into account. Figure \ref{fig:flow} shows a schematic diagram of the proposed two-stage load forecasting approach aimed at improving load forecasts. To enhance the computational efficiency of the proposed two-stage approach, we further propose to perform input-space dimensionality reduction via a model-free global sensitivity analysis (GSA) method for input feature selection. 

The proposed two-stage approach is tested on various IEEE distribution system test cases, namely, 8-bus, 14-bus, 24-bus and 123-bus systems, in comparison with representative one-stage and two-stage methods. The proposed two-stage approach outperforms the competing methods in terms of both point and interval predictions across all cases tested.

The remainder of the paper is structured as follows. 
Section \ref{sec:two-stage} elaborates on the proposed two-stage approach. Section \ref{sec:selection} presents the GSA method that aims at enhancing the computational efficiency. Section \ref{sec:numerical} presents numerical experiments for comparing the proposed two-stage approach with competing forecasting methods. Section \ref{sec:conclusion} concludes the paper.

\section{Load Forecasting with Input Modeling and Uncertainty Quantification}\label{sec:two-stage}

In this section, we elaborate on how the proposed two-stage approach can thoroughly address the issue of input errors by quantifying and propagating their impact into the ultimate load forecasts.

For ease of exposition, we first define the input features to be used in the second-stage model.  Consider predicting customer $i$'s load at hour $t$ in a distribution grid comprising $N$ customers. We adopt the following set of input features:
\begin{eqnarray}\label{eqn:input}
\mathbf{x}_t=\left(\theta_{i,1}^t, \ldots, \theta_{i,i-1}^t, \theta_{i,i+1}^t, \theta_{i,i+2}^t, \ldots,\theta_{i,N}^t\right)^\top,
\end{eqnarray}
 with $\theta_{i,k}$ denoting the difference in phase angles of customers $i$ and $k$. Such a choice was first adopted by the IGP framework proposed in \cite{Xie}, which is inspired by the well-known power flow equation:
\begin{eqnarray} \label{eqn:power}
0=-P_i+\sum\limits_{k=1}^N|V_i||V_k|(G_{ik}\cos\theta_{ik}+B_{ik}\sin\theta_{ik}),
\end{eqnarray}
where $P_i$ denotes the load of bus $i$ in a distribution grid, $|V_i|$ is the voltage magnitude of bus $i$, and $G_{ik}$ and $B_{ik}$ are coefficients describing the relationship between buses $i$ and $k$, with $i$, $k=1,2,\ldots,N$. Equation (\ref{eqn:power}) implies that a given target customer's load is strongly related to the phase angle differences between other customers and herself.

It is known in the literature that there typically exists a strong coupling between load and phase angle, whereas the coupling between load and voltage magnitude is rather weak \cite{Monticelli2006}. Hence, the use of phase angle information as inputs is arguably sufficient for load forecasting purposes. We note that other relevant input features such as weather conditions and geographic locations can be easily incorporated into the proposed two-stage approach as well. For simplicity, we keep the input features as given in (\ref{eqn:input}) in this work.

\subsection{Input Modeling and Error Quantification in the First Stage}\label{sec:firststage}

\subsubsection{Model Choice}\label{sec:nngp}
Consider predicting target customer $i$'s hour-ahead load $P_i^{t+1}$ at hour $t$, the second-stage model in our two-stage approach requires an input vector $\mathbf{x}_{t+1}$ in the form of (\ref{eqn:input}) for load prediction. Since the input values in $\mathbf{x}_{t+1}$ are unknown to us at hour $t$, the first-stage model is adopted to perform hour-ahead input prediction. To fully quantify the input prediction errors, the first-stage model must provide an interval prediction as well as a point prediction. GP becomes a natural choice to this end thanks to its inherent predictive uncertainty measure. However, GP is known for its incompetent extrapolation performance \cite{phdthesis}, hence unsuitable as the first-stage model for predicting future input values. In contrast, NNs are known for their ability to capture complex output patterns and their outstanding extrapolation performance. However, NNs do not come with a predictive uncertainty measure. We propose to use the neural network--Gaussian process (NNGP) model recently proposed in \cite{DNN} to accomplish both tasks simultaneously. The underlying idea of NNGP is to place prior distributions on the weight and bias parameters of a deep neural network (DNN) with infinite width and derive its equivalent GP formulation. By using the equivalent GP for prediction, NNGP can combine the strengths of NN and GP, providing a highly accurate point prediction as well as an informative interval prediction for future input values.

Below we briefly explain the rationale behind NNGP. Consider a fully-connected DNN with $L$ hidden layers ($L\geq 2$), where the number of hidden units in each layer, $N_u$, are equal. Denote the pointwise nonlinearity function by $\phi$. Let $\mathbf{x}\in \mathbb{R}^{d_{in}}$ and $\mathbf{y}\in \mathbb{R}^{d_{out}}$ denote the input and output vectors of the network, respectively. Consider the $i$th hidden unit in the $l$th layer, and denote the post-nonlinearity and the post-affine transformation by $a_i^l$ and $z_i^l$, respectively. Hence, we have $\mathbf{y}=\left(z_1^L(\mathbf{x}), z_2^L(\mathbf{x}), \ldots, z_{d_{out}}^L(\mathbf{x})\right)^\top$. Denote the weight and bias parameters in the $l$th layer by $W_{ij}^l$ and $b_i^l$; we place normal prior distributions with zero means and variances $\sigma_{w}^2/N_u$ and $\sigma^2_b$ respectively on $W_{ij}^l$ and $b_i^l$. 

To derive an equivalent GP for a generic DNN at hand, we first show how to do so for a one-layer NN with infinite width. Consider the output of the $i$th hidden unit of the first layer in the NN, we have 
\begin{eqnarray}
z_i^1(\mathbf{x}) &=& b_i^1+\sum\limits_{j=1}^{N_u}W_{ij}^1a_j^1(\mathbf{x}), \nonumber\\ 
\text{with} \ \ a_j^1(\mathbf{x}) &=& \phi\left(b_j^0+\sum\limits_{k=1}^{d_{in}} W^0_{jk}x_k\right), \nonumber
\end{eqnarray}
where the superscript ``0'' denotes the input layer of the NN, and $x_k$ denotes the $k$th component of the input vector $\mathbf{x}$. As $z_i^1(\mathbf{x})$ is the sum of independent and identically distributed terms, the standard Central Limit Theorem indicates that $z_i^1(\mathbf{x})$ becomes normally distributed as $N_u \rightarrow \infty$. Hence, given any input vectors $\mathbf{x}^1$, $\mathbf{x}^2$, \ldots, $\mathbf{x}^n$, the vector $\mathbf{z}_i^1=\left(z_i^1(\mathbf{x}^1), z_i^1(\mathbf{x}^2), \ldots, z_i^1(\mathbf{x}^n)\right)^\top$ follows a multivariate normal distribution. That is, $\mathbf{z}_i^1\sim \mathcal{N}(\mu^1,K^1)$, where $K^1$ denotes the $n \times n$ covariance matrix for the first layer, with its $(g,h)$th component given by  $K^1(g,h)=\Cov(z_i^1(\mathbf{x}^g),z_i^1(\mathbf{x}^h))$, $g,h=1,2,\ldots,n$. Since the parameters $W_{ij}^1$'s and $b_i^1$'s have zero means, we have $\mu^1(\mathbf{x})=\mathbb{E}[z_i^1(\mathbf{x})]=0$ for any $\mathbf{x}\in \mathbb{R}^{d_{in}}$. In particular, given any two inputs $\mathbf{x}$, $\mathbf{x}'\in \mathbb{R}^{d_{in}}$, it follows that
\begin{eqnarray} \label{eqn:cov_first}
K^1(\mathbf{x},\mathbf{x}')=\mathbb{E}[z_i^1(\mathbf{x})z_i^1(\mathbf{x}')]=\sigma^2_b+\sigma^2_w\mathbb{E}[a_i^1(\mathbf{x})a_i^1(\mathbf{x}')].
\end{eqnarray}
We can see from the above that each component of the output vector of a one-layer NN with infinite width is equivalent to a GP with a covariance function specified by (\ref{eqn:cov_first}). 

The aforementioned result can be generalized to NNs with more than one layer in a recursive manner by utilizing the relationship between $a_j^l$ and $z_j^{l-1}$, i.e. $a_j^l(\mathbf{x})=\phi(z_j^{l-1}(\mathbf{x}))$. The resulting covariance function in the $L$th layer, $K^L(\mathbf{x},\mathbf{x}')$, hence serves as the covariance kernel of the equivalent GP for a DNN with $L$ hidden layers. For the sake of brevity, we refer the interested reader to Appendix \ref{sec:app_NNGP} for a detailed derivation.

\subsubsection{Input Prediction and Error
Quantification}\label{sec:usenngp}
In the first stage, we predict the hour-ahead input vector component by component via NNGP. For predicting each component, we utilize not only their own values observed in the past but also the observed values of other components. 

Let us consider predicting the input $\theta_{i,j}$ at hour $h^*$, for $j=1, 2,  \ldots,i-1,i+1, \ldots, N$ (see the definition given in \eqref{eqn:input}). We construct a training set for NNGP as $\{D_j^{h^*-24n_{t_1}},D_j^{h^*-24n_{t_1}+1},\ldots,D_j^{h^*-1}\}$, where $D_j^{h}$ denotes an input-output pair $(X_j^h, Y_j^h)$ for $h=h^*-24n_{t_1}, h^*-24n_{t_1}+1, \ldots, h^*-1$, and $n_{t_1}$ denotes the number of days from which the training points are sampled. For our purpose, the output layer $Y_j^h$ is in fact $\theta_{i,j}^{h}$, namely, the phase angle difference between customer $i$ and customer $j$ at hour $h$. Taking into account the correlations between different customers (recall \eqref{eqn:input}, \eqref{eqn:power} and discussions therein), we set $X_{j}^h=(\Theta_{i,j}^{h\top},\Theta_{i,j^1}^{h\top},\ldots,\Theta_{i,j^{n_j}}^{h\top})^\top$, where $\Theta_{i,j}^h=(\theta_{i,j}^{h-24n_{in}},\ldots,\theta_{i,j}^{h-1})^\top$, with $n_{in}$ denoting the number of days from which the data in the input layer are sampled,  and the indices $j^{1},j^{2},\ldots,j^{n_j}$ are selected from $\{1,2,\ldots,N\}$ excluding $i$ and $j$. 

Upon constructing the training set, NNGP can be trained using the covariance kernel obtained following the steps as given in Section \ref{sec:nngp}. The predictive mean and variance of each component in the hour-ahead input vector can be obtained via NNGP in the same manner as using a GP model. Hence, the predictive mean of the input vector
\begin{eqnarray}\label{eqn:mean_input}
\mu_1=(\mathbb{E}[\theta_{i,1}],\ldots,\mathbb{E}[\theta_{i,i-1}],\mathbb{E}[\theta_{i,i+1}],\ldots,\mathbb{E}[\theta_{i,N}])^\top
\end{eqnarray}
and predictive variance of the input vector 
\begin{eqnarray}\label{eqn:variance_input}
\mathsf{V}_1= (\mathbb{V}[\theta_{i,1}],\ldots,\mathbb{V}[\theta_{i,i-1}],\mathbb{V}[\theta_{i,i+1}],\ldots,\mathbb{V}[\theta_{i,N}])^\top
\end{eqnarray}
are readily available, and can then be used in the second-stage model for making ultimate load forecasts.

The choice of the parameters for NNGP (i.e., $n_{t_1}$, $n_{in}$, $\sigma_b$, $\sigma_w$, and the number of hidden layers $L$) greatly impacts the predictive accuracy achieved. Appropriate parameter values can be obtained by a cross-validation procedure, and we refer the interested reader to \cite{DNN} for details. 

\subsection{Input Error Propagation and Uncertainty Quantification in the Second-Stage Load Forecasting}\label{sec:secondstage}

\subsubsection{Model Choice}\label{sec:review-NIGP}
In the second stage, we aim to appropriately propagate the impact of input errors assessed in the first stage into the ultimate load forecasting results, reflecting it in both point and interval predictions. The noisy input Gaussian process (NIGP) model proposed in \cite{McHutchon2011} is arguably an adequate choice as the second-stage model. Compared to standard GP models, NIGP can propagate the impact of input errors into both point and interval predictions, providing a thorough solution to the issue of input errors in load forecasting.

Below we briefly describe the rationale behind NIGP.
Assume that the observed output $y\in \mathbb{R}$ is a noisy measurement of an actual output $\tilde{y}$, that is,
\begin{eqnarray}\label{eqn:y}
y=\tilde{y}+\varepsilon_y,
\end{eqnarray}
where the output noise is assumed to be normally distributed, i.e., $\varepsilon_y \sim \mathcal{N}(0,\sigma^2_y)$, with $\sigma^2_y$ denoting the output noise variance. Assume that the observed input vector $\mathbf{x}\in \mathbb{R}^d$ can be modeled as the actual input vector $\tilde{\mathbf{x}}$ corrupted by some input noise:
\begin{eqnarray}\label{eqn:x}
\mathbf{x}=\tilde{\mathbf{x}}+\bm{\varepsilon}_x,
\end{eqnarray}
where the input noise vector is assumed to be normally distributed, i.e., $\bm{\varepsilon}_x \sim \mathcal{N}(0,\Sigma_x)$, with $\Sigma_x$ denoting the $d\times d$ input noise covariance matrix. Based on (\ref{eqn:y}) and (\ref{eqn:x}), we can establish the following relationship between the observed input-output pair $(\mathbf{x},y)$ as follows:
\begin{eqnarray} 
\label{eq:noisy-output-model}
y=f(\mathbf{x}-\bm{\varepsilon}_x)+\varepsilon_y,
\end{eqnarray}
where $f: \mathbb{R}^d \rightarrow \mathbb{R}$ denotes the target input-output relationship to estimate. Suppose that $f(\cdot)$ can be modeled as a GP with some covariance kernel. The first-order Taylor expansion of $f(\mathbf{x}-\bm{\varepsilon}_x)$ yields
\begin{eqnarray}\label{eqn:taylor}
f(\mathbf{x}-\bm{\varepsilon}_x)=f(\mathbf{x})-\bm{\varepsilon}^\top_x \frac{\partial{f(\mathbf{x})}}{\partial{\mathbf{x}}}+\ldots,
\end{eqnarray}
where $\partial{f(\cdot)}/\partial{\mathbf{x}}$ denotes the partial derivative process of $f(\cdot)$, which is also a GP \cite{Ras2006GP}. Given that the distribution of the product of two normal random vectors $\bm{\varepsilon}_x$ and $\partial{f(\mathbf{x})}/\partial{\mathbf{x}}$ is analytically intractable, we approximate $\partial{f(\mathbf{x})}/\partial{\mathbf{x}}$ by the gradient of the posterior mean of $f(\mathbf{x})$, and denote it by  $\bm{\partial}\mathbf{\bar{f}}$. It then follows from \eqref{eq:noisy-output-model} and \eqref{eqn:taylor} that
\begin{eqnarray}\label{eqn:approx_y}
y \approx f(\mathbf{x})-\bm{\varepsilon}_x^\top\bm{\partial}\mathbf{\bar{f}}+\varepsilon_y.
\end{eqnarray}
We see from (\ref{eqn:approx_y}) that the observed output $y$ is approximately normally distributed, i.e., 
$P(y|f)=\mathcal{N}(f,\sigma_y^2+\bm{\partial}\mathbf{\bar{f}}^\top\Sigma_x\bm{\partial}\mathbf{\bar{f}})$.

Now suppose that the training set for NIGP comprises $N$ observed input-output pairs, $\{(\mathbf{x}_i, y_i)\}_{i=1}^N$. Denote the $N\times d$ input matrix by  $\mathbf{X}$ and the $N$-dimensional output vector by $Y$. Let $\mathbf{C}$ denote the $N\times N$ covariance matrix obtained by evaluating the covariance kernel at $\mathbf{X}$, and denote the gradient of the posterior means at the $N$ training inputs by $\bm{\Delta}\mathbf{\bar{f}}$, which is an $N\times d$ matrix. When the popular squared exponential covariance kernel is used, the covariance between any two inputs $\mathbf{x}_i$ and $\mathbf{x}_j$ is given by 
\begin{eqnarray}\label{eq:spatial-cov-NIGP}
\mathbf{C}(\mathbf{x}_i,\mathbf{x}_j)=\sigma_f^2\exp\big(-\frac{1}{2}(\mathbf{x}_i-\mathbf{x}_j)^\top\Lambda^{-1}(\mathbf{x}_i-\mathbf{x}_j)\big),  
\end{eqnarray}
where $\sigma_f^2$ represents the process variance, and $\Lambda$ denotes the $d\times d$ diagonal matrix with its main diagonal elements being the length-scale hyper-parameters. 

The expressions for the predictive mean and variance of NIGP can be derived when the test input is deterministic and stochastic. Denote the observed test input by $\mathbf{x}_*$ and the actual test input by $\tilde{\mathbf{x}}_*$. For the deterministic case, we have $\mathbf{x}_*=\tilde{\mathbf{x}}_*$. Hence,
\begin{eqnarray}
\mathbb{E}[f(\mathbf{x}_*)|\mathbf{X},Y]&=&\mathbf{C}(\tilde{\mathbf{x}}_*,\mathbf{X})[\mathbf{C}(\mathbf{X},\mathbf{X})+\sigma^2_y \mathbf{I}_N \nonumber \\ && +
\diag(\mathbf{\Delta\bar{f}}\Sigma_x \mathbf{\Delta\bar{f}}^\top)]^{-1}Y, \label{gpmean_det}  \\
\mathbb{V}[f(\mathbf{x}_*)|\mathbf{X},Y]&=&\mathbf{C}(\tilde{\mathbf{x}}_*,\tilde{\mathbf{x}}_*)-\mathbf{C}(\tilde{\mathbf{x}}_*,\mathbf{X})[\mathbf{C}(\mathbf{X},\mathbf{X})+\sigma^2_y\mathbf{I}_N\nonumber\\ && +\diag(\mathbf{\Delta\bar{f}}\Sigma_x \mathbf{\Delta\bar{f}}^\top)]^{-1}\mathbf{C}(\mathbf{X},\tilde{\mathbf{x}}_*), \nonumber
 \end{eqnarray}
where $\diag(\mathbf{\Delta\bar{f}}\Sigma_x \mathbf{\Delta\bar{f}}^\top)$ denotes the diagonal of the matrix $\mathbf{\Delta\bar{f}}\Sigma_x \mathbf{\Delta\bar{f}}^\top$, and $\mathbf{I}_N$ denotes the $N\times N$ identity matrix. Notice that $\mathbf{\Delta\bar{f}}\Sigma_x \mathbf{\Delta\bar{f}}^\top$ is a correction term added to the covariance matrix of a standard GP model to reflect the uncertainty in the inputs $\mathbf{X}$.

In the stochastic case, assume that the observed test input $\mathbf{x}_*$ is multivariate normally distributed, i.e., $\mathbf{x}_* \sim \mathcal{N}(\tilde{\mathbf{x}}_*,\Sigma_x)$. 
It follows that the predictive mean of $f(\mathbf{x}_*)$ can be given as
\begin{eqnarray}\label{eqn:mean}
\mathbb{E}[f(\mathbf{x}_*)]=\left([\mathbf{C}+\sigma^2_y \mathbf{I}_N+\diag(\mathbf{\Delta\bar{f}}\Sigma_x \mathbf{\Delta\bar{f}}^\top)]^{-1}Y \right)^\top \mathbf{q}. 
\end{eqnarray}
Compared to the predictive mean given in (\ref{gpmean_det}), $\mathbf{q}$ is the counterpart of $\mathbf{C}(\tilde{\mathbf{x}}_*,\mathbf{X})$ in the stochastic case, which is given by
\begin{eqnarray}
\mathbf{q}=\int \mathbf{C}(\tilde{\mathbf{x}}_*, \mathbf{X})p(\tilde{\mathbf{x}}_*|\mathbf{x}_*,\Sigma_x) d \tilde{\mathbf{x}}_*, \nonumber
\end{eqnarray}
where $p(\tilde{\mathbf{x}}_*|\mathbf{x}_*,\Sigma_x)$ is the posterior of $\tilde{\mathbf{x}}_*$. When the squared exponential covariance kernel in (\ref{eq:spatial-cov-NIGP}) is adopted, $\mathbf{q}$ is an $N\times 1$ vector whose $i$th component is given by
\begin{eqnarray*}
q_i &= & \sigma_f^2|\Sigma_x\Lambda^{-1}+\mathbf{I}_d|^{-\frac{1}{2}}\exp(-\frac{1}{2}(\mathbf{x}_i-\tilde{\mathbf{x}}_*)^\top (\Sigma_x+\Lambda)^{-1} \nonumber \\
&& (\mathbf{x}_i-\tilde{\mathbf{x}}_*)), \quad  i=1,2,\ldots,N,
\end{eqnarray*}
where $\mathbf{I}_d$ denotes the $d\times d$ identity matrix. Utilizing the law of total variance, the predictive variance of $f(\mathbf{x}_*)$ follows as
\begin{eqnarray} \label{eqn:var}
\mathbb{V}[f(\mathbf{x}_*)] & = & \sigma_f^2 +  \bm{\alpha}^\top \mathbf{Q} \bm{\alpha}- (\mathbb{E}[f(\mathbf{x}_*)])^2   \\
&-&\tr([\mathbf{C}+\sigma^2_y \mathbf{I}_N+\diag(\mathbf{\Delta\bar{f}}\Sigma_x \mathbf{\Delta\bar{f}}^\top)]^{-1}\mathbf{Q}) \nonumber
\end{eqnarray}
where $\tr(\bm{M})$ denotes the trace of matrix $\bm{M}$, and $\mathbf{Q}$ is an $N\times N$ matrix whose $(i,j)$th component is given by
\begin{eqnarray*}
\mathbf{Q}_{ij} &= & \frac{\mathbf{C}(\mathbf{x}_i,\tilde{\mathbf{x}}_*)\mathbf{C}(\mathbf{x}_j,\tilde{\mathbf{x}}_*)}{|2\Sigma_x\Lambda^{-1}+\mathbf{I}_d|^\frac{1}{2}}\exp((\mathbf{z}-\tilde{\mathbf{x}}_*)^\top (\Lambda +\frac{1}{2}\Lambda \Sigma_x^{-1} \nonumber \\
&& \Lambda)^{-1}(\mathbf{z}-\tilde{\mathbf{x}}_*)), \ \mbox{with} \  \mathbf{z}=(\mathbf{x}_i+\mathbf{x}_j)/2,
\end{eqnarray*}
and 
$\bm{\alpha}=(\mathbf{C}+\sigma^2_y \mathbf{I}_N+\diag(\mathbf{\Delta\bar{f}}\Sigma_x \mathbf{\Delta\bar{f}}^\top))^{-1}Y.$
We refer the interested reader to \cite{McHutchon2011,phdthesis} for more details on the derivation of the predictive mean and variance functions of NIGP. A MATLAB package is available for practical implementation of NIGP \cite{NIGPcode}.

\subsubsection{Input Error Propagation and Quantification in Load Forecasting}
In the second stage, NIGP can be conveniently implemented for making ultimate load forecasts. As NIGP is essentially a variant of GP, its training and prediction can be performed in the same manner as using a standard GP. Consider predicting the target customer $i$'s hour-ahead load $P_i^{h^*}$ at hour $h^*-1$. We first estimate the hyper-parameters $\sigma^2_f$, $\Lambda$ and $\sigma^2_y$ via maximum likelihood estimation with a set of training input-output pairs $\{(\mathbf{x}_h,P_i^h)\}_{h=1}^{n_{t_2}}$, where $n_{t_2}$ denotes the number of days from which the training points are sampled, and $\mathbf{x}_h$ is in the form of \eqref{eqn:input}. With the estimates of the hyper-parameters, the load forecast $\widehat{P}_i^{h_*}$ can be made with the predictive mean and variance of the input vector, $\mu_1$ and $\mathsf{V}_1$, respectively given by (\ref{eqn:mean_input}) and (\ref{eqn:variance_input}). Specifically, set $\Sigma_1= \diag(\mathsf{V}_1)$, a diagonal matrix whose main diagonal entries are given by the components of $\mathsf{V}_1$. Then the predictive mean and variance for the target customer's hour-ahead load $P_i^{h^*}$ can be obtained by replacing $\tilde{\mathbf{x}}_*$ and $\Sigma_{x}$ in (\ref{eqn:mean}) and (\ref{eqn:var}) respectively by $\mu_1$ and $\Sigma_1$. The predictive mean can be used as a point prediction for $P_i^{h^*}$, and the prediction variance can be utilized to construct an interval prediction for $P_i^{h^*}$.

\section{Improving Computational Efficiency via Global Sensitivity Analysis}\label{sec:selection}

In this section, we aim to improve the computational efficiency of the proposed two-stage load forecasting approach. Due to the use of spatial information (i.e. phase angle difference) for load forecasting, the input-space dimensionality can increase dramatically with the number of customers in the power grid, resulting in a high computational cost. To reduce the input-space dimensionality, one can consider retaining only a few most important input dimensions for load forecasting. Although many feature selection techniques (e.g., filter methods, wrapper methods, and embedded methods, see \cite{Guyon2003,Kumar2014}) are available, most of them are parametric approaches, hence pose restrictive assumptions on the underlying input-output relationship. In this work, we adopt the variance-based GSA method, which is model-free. GSA can be applied to both stages of our two-stage approach to improve the computational efficiency achieved.

\subsection{Variance-Based GSA for Systems with Functional Inputs }\label{sec:SA}

GSA focuses on quantifying how sensitive the output is to each individual input feature and their interactions, and identifies those inputs that contribute the most to the output variability. GSA methods fall into two categories: regression-based and variance-based. The main idea of the variance-based GSA methods is to decompose the variance of the output as a sum of contributions of each
input feature. One of the most widely used variance-based GSA methods is the Sobol' index method, on which we provide a brief overview below.

Consider the following model: $X \mapsto Y = f(X)$, where $f(\cdot)$ denotes the underlying input-output function, $X=(X_1,X_2\ldots, X_m)^\top$ denotes the $m\times 1$ vector of input features, and $Y$ represents the output.
The first-order Sobol' index of $X_i$ is defined as 
\begin{eqnarray*}
S_i = \frac{\mathbb{V}_{X_i}(\mathbb{E}_{X_{\sim i}}[Y|X_i])}{\mathbb{V}(Y)},
\end{eqnarray*}
which quantifies the impact of $X_i$ on $Y$ only, excluding any interactions with other inputs. The total Sobol' index of $X_i$ is defined as 
\begin{eqnarray*}
S_{T_i} = \frac{\mathbb{E}_{X_{\sim i}}[\mathbb{V}_{X_i}(Y|X_{\sim i})]}{\mathbb{V}(Y)}=1-\frac{\mathbb{V}_{X_{\sim i}}(\mathbb{E}_{X_i}[Y|X_{\sim i}])}{\mathbb{V}(Y)},
\end{eqnarray*}
where $X_{\sim i}$ denotes the collection of all inputs except $X_i$. The total Sobol' index measures the total impact of $X_i$ on $Y$, including higher-order impacts through interactions with other inputs. By definition, a Sobol' index takes a value in $[0,1]$. The larger an index value is, the greater impact the associated input has on the output. 

A majority of existing methods proposed for Sobol' index estimation focus on tackling spatial inputs, see, e.g., \cite{Saltelli2010,Marrel2011,Kucherenko2012,Norton2015}. In our problem context, however, methods capable of handling functional inputs (i.e., inputs that vary with time) are required. GSA methods that tackle systems with functional inputs are relatively scarce.
In \cite{Nanty2016}, the authors proposed an effective Sobol' index estimation method for GSA of dynamic systems, which is suitable to be applied to the first- and second-stage analyses to improve the computational efficiency achieved.  

To estimate the Sobol' indices, we first decompose the functional inputs through a simultaneous principle component analysis as follows:
 \begin{eqnarray} 
X=\sum\limits_{i=1}^p \beta_i\gamma_i,
\label{eqn:SPCA}
\end{eqnarray}
 where $\bm{\gamma}=\{\gamma_1,\gamma_2,\ldots,\gamma_p\}$ denotes the collection of basis functions such as B-splines and wavelets, and $\bm{\beta}= \{\beta_1,\beta_2\ldots,\beta_p\}$ denotes the set of corresponding coefficients, with $p$ denoting the number of basis functions used. Then we approximate the joint distribution of the coefficients in $\bm{\beta}$ using a Gaussian mixture model. Subsequently, we can construct Monte Carlo (MC) samples of the functional inputs based on (\ref{eqn:SPCA}) by sampling from the approximated distribution of $\bm{\beta}$. Given a sample of input features, we can obtain the corresponding outputs $Y$ using an approximated input-output function $\widehat{f}$. The first-order and total Sobol' indices can then be estimated via MC sampling. Taking into account the potential interactions between the components of the input vector, we adopt the total Sobol' index as the importance measure and use the following estimator proposed in \cite{Jansen1999}: 
 \begin{eqnarray}\label{eq:total-index-estimate} 
\widehat{S}_{T_i}=\frac{{(2n)}^{-1}\sum\limits_{j=1}^n\left(\widehat{f}(\mathbf{A})_j-\widehat{f}(\mathbf{A}_\mathbf{B}^{(i)})_j\right)^2}{\mathbb{V}(\widehat{f}(\mathbf{C}))},
\end{eqnarray}
where $\mathbf{A}$, $\mathbf{B}$, and $\mathbf{C}$ respectively denote three independently generated $n\times m$ matrices of input samples, each row of which gives a randomly sampled input vector. In \eqref{eq:total-index-estimate}, $\widehat{f}(\mathbf{A})$ evaluates the approximated function $\widehat{f}$ at the input vector given by each row of matrix $\mathbf{A}$, and the subscript $j$ of $\widehat{f}(\mathbf{A})_j$ denotes the $j$th row of the resulting matrix $\widehat{f}(\mathbf{A})$. $\mathbf{A}_\mathbf{B}^{(i)}$ denotes the matrix whose $i$th column is taken from matrix $\mathbf{B}$ while the other
$m-1$ columns remain from matrix $\mathbf{A}$.


\subsection{Speeding Up Input Prediction and Uncertainty
Quantification}\label{sec:firstSA}
The purpose of performing GSA in the first stage is to select a set of customer indices $\{j^1,j^2,\ldots,j^{n_j}\}$ from $\{1,2,\ldots,N\}$ (see Section \ref{sec:usenngp}) to help with input prediction. Consider predicting the value of input $\theta_{i,j}$ at each hour on a particular day, namely, $\theta_{i,j}^{h^*},\ldots,\theta_{i,j}^{h^*+23}$. Prior to prediction, we first construct a training set with full-dimensional inputs for the GSA purpose. Specifically, we set $n_j=N-2$ and $n_{in}=1$ in Section \ref{sec:usenngp}. Hence, each input dimension corresponds to the phase angle difference of the target customer and another customer at one of the previous 24 hours and the input-space dimensionality is $24(N-1)$. Recall that in the first stage, the input-output relationship is approximated by NNGP, hence the first-stage GSA is carried out based on the NNGP model to obtain the total index estimates, $\widehat{S}^1_{T_i}$, $i=1,2,\ldots,24(N-1)$. Since each index $i$ corresponds to a particular hour and a particular customer, we then quantify the impact of each component of the input vector $(\theta_{i,1},\ldots,\theta_{i,i-1},\theta_{i,i+1},\ldots,\theta_{i,N})^\top$ by the following summary statistic:
\begin{eqnarray*}
H_{\ell}=\sum\limits_{i=24(\ell-1)+1}^{24\ell} \widehat{S}^1_{T_i}, \quad  \ell=1,2,\ldots,N-1. 
\end{eqnarray*}
Here $H_\ell$ measures the impact of each input feature on the output in the first-stage input prediction, and the larger the value, the greater the impact. We sort the $H_\ell$'s in a non-increasing order and select the first $n_j$ corresponding components (excluding $\Theta_{i,j}^h$) to be included in $X_{j}^h$ for input prediction as detailed in Section \ref{sec:usenngp}. The value of $n_j$ can be determined together with other parameters of NNGP by running a cross-validation procedure.  

\subsection{Speeding Up Load Forecasting}\label{sec:secondSA}
In the seccond stage, we aim to identify and select a few components in the input vector $(\theta_{i,1}, \ldots, \theta_{i,i-1}, \theta_{i,i+1}, \ldots, \theta_{i,N})^\top$ to be used in the NIGP model for predicting the target customer $i$'s hour-ahead load. 
We carry out GSA based on the NIGP model trained in the second stage following the steps detailed in Section \ref{sec:SA}. We obtain the total Sobol' index estimate, $\widehat{S}_{T_j}^2$, for each input $\theta_{i,j}$. The higher $\widehat{S}_{T_j}^2$ is, the more important $\theta_{i,j}$ is considered for load prediction of the target customer $i$. Upon identifying the most important components in the input vector, only those components will be retained as inputs for future load prediction. Hence, a higher computational efficiency can be achieved. 

\section{Numerical Experiments}\label{sec:numerical}
In this section, we numerically evaluate the predictive performance of the proposed two-stage load forecasting approach against two benchmark methods, i.e., a representative two-stage model --- IGP proposed in \cite{Xie} and a representative one-stage model --- SARIMA studied in \cite{Pappas2008}. We refer to the proposed two-stage approach as NNGP-NIGP. As mentioned in Section \ref{sec:two-stage}, IGP adopts the same input features as NNGP-NIGP. Nonetheless, the first stage of IGP only provides a point prediction for the hour-ahead input vector via $k$-means clustering, and fails to quantify and propagate the impact of input errors into ultimate load forecasts. Without a mechanism to properly tackle the uncertain inputs, IGP constructs interval predictions for the target customer's load via a heuristic approach. As will be shown later on, the comparison of NNGP-NIGP and IGP clearly manifests the substantial improvement in the predictive performance brought about by proper input-error propagation and quantification. The SARIMA model, on the other hand, modifies the traditional seasonal ARIMA model with a multi-model partitioning algorithm, and uses the target customer's data for forecasting. This one-stage model serves as a baseline for comparison in the numerical experiments. 
 
\subsection{Experimental Setup}\label{sec:exp_setup}
Experiments are conducted on four IEEE test cases, respectively, 8-bus, 14-bus, 123-bus test cases as well as a 24-bus test case which is constructed to simulate the phase unbalance problem based on the 8-bus test case. To simulate highly uncertain load behaviors caused by DERs in real-life power systems, historical load profiles drawn from real-world data sources are used for simulations. Taking into account the uncertain renewable generation behaviors of DERs, 
we first pre-process the hourly PV generation data over a year drawn from Renewable.ninja \cite{Renewable}, and then subtract the pre-processed data from the load data of each bus. To obtain phase angle values, we perform power flow analysis to generate the states of the power system hourly over a one-year time frame using the MATLAB Power
System Simulation Package (MATPOWER) \cite{Zimmerman2009Mat,Zimmerman2010Mat}, based on the processed load data.

The predictive performance of each approach is evaluated by the accuracy of the point prediction and the coverage ability of the predictive interval obtained. To evaluate point predictive accuracy, we use the mean absolute percentage error (MAPE), which is defined below:
\begin{eqnarray}
\text{MAPE}={T}^{-1}\sum\limits_{h=1}^{T}\left|\frac{P_h-\widehat{P}_h}{P_h}\right|, \nonumber
\end{eqnarray}
where $P_h$ and $\widehat{P}_h$ respectively denote the load actually observed and the predicted load at hour $h$, and $T$ denotes the number of hourly predictions made. In our experiments, $T=24$ is used to calculate MAPE for every prediction day. Since the SARIMA model does not provide an interval prediction, we only compare the coverage abilities of the predictive intervals given by NNGP-NIGP and IGP.
The coverage abilities are assessed by the coverage probability (CP):
\begin{eqnarray*}
\text{CP}= n_c/24, \nonumber
\end{eqnarray*}
where $n_c$ denotes the number of observed load data points covered by the predictive interval on a prediction day.
Predictions are performed over 30 days in each season for all customers in each test case considered. 

To set appropriate parameter values for the first-stage NNGP model and GSA, a cross-validation procedure was adopted. For the 8-bus, 14-bus, and 24-bus test cases, the parameter setting adopted is $n_t=60$, $n_{in}=3$, $\sigma_b=1$, $\sigma_w=1$, and $n_j=1$  (the number of neighboring customers whose information is included for the first-stage input prediction; see Section \ref{sec:firstSA} for details). For the 123-bus test case, the parameter setting used is $n_t=90$, $n_{in}=3$, $\sigma_b=1$, $\sigma_w=1$, and $n_j=2$. Regarding the second-stage NIGP model, the only parameters that require estimation are those in the covariance kernel, which are learned via maximum likelihood estimation. The numerical experiments are performed on a laptop with 6th generation Intel\textsuperscript{\circledR} \ Core\textsuperscript{TM} i7 processor and 8.0GB DDR4 memory. 

\subsection{Comparison of Point and Interval Predictions}\label{sec:point_interval}
\begin{figure}[ht]
  \centering%
  \includegraphics[width=8cm]{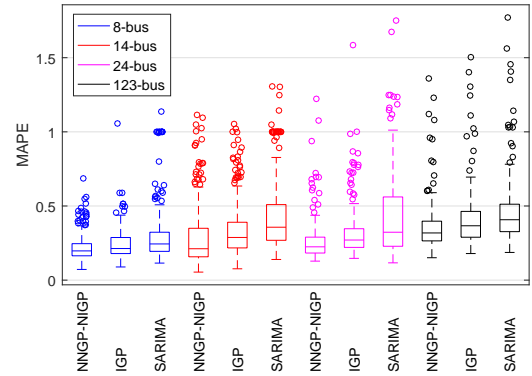}
  \caption{Boxplots of MAPEs obtained by NNGP-NIGP, IGP, and SARIMA.}
  \label{fig:box_mape}
\end{figure}

The MAPEs obtained by NNGP-NIGP, IGP, and SARIMA in all test cases are summarized in Fig. \ref{fig:box_mape}. We see that across all test cases, NNGP-NIGP outperforms IGP and SARIMA in terms of the point estimation accuracy. This is because NNGP-NIGP can properly incorporate the impact of input errors in the ultimate point prediction for the target customer's load, while the competing methods fail to do so.  
 
\begin{figure}[ht]
  \centering%
  \includegraphics[width=8cm]{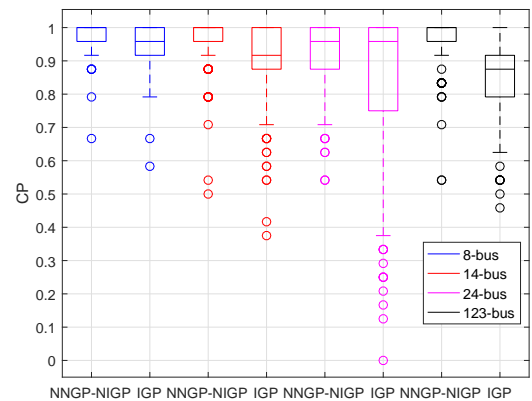}
  \caption{Boxplots of coverage probabilities obtained by NNGP-NIGP and IGP.}
  \label{fig:box_cov_8bus}
\end{figure}

\begin{figure}
  \centering%
  \subfloat[8-bus test case.]{\label{fig:pred_8}\includegraphics[width=6.5cm]{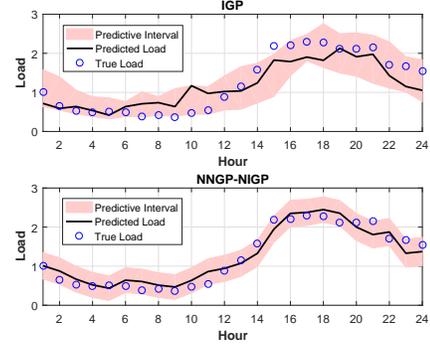}}\\
  \subfloat[14-bus test case.]{\label{fig:pred_14}\includegraphics[width=6.5cm]{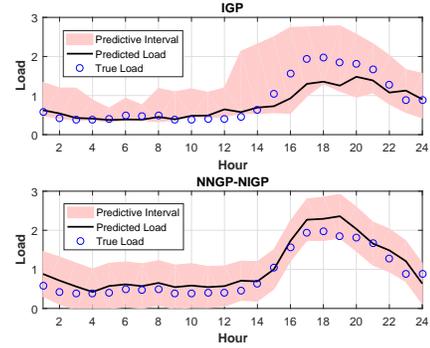}}\\
   \subfloat[24-bus test case.]{\label{fig:pred_24}\includegraphics[width=6.5cm]{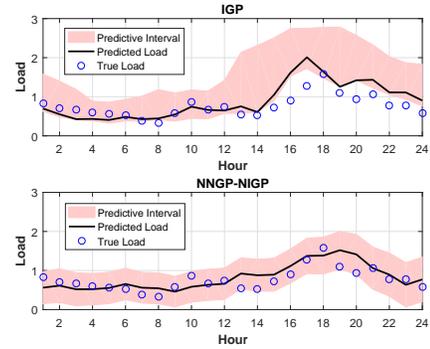}}\\
   \subfloat[123-bus test case.]{\label{fig:pred_123}\includegraphics[width=6.5cm]{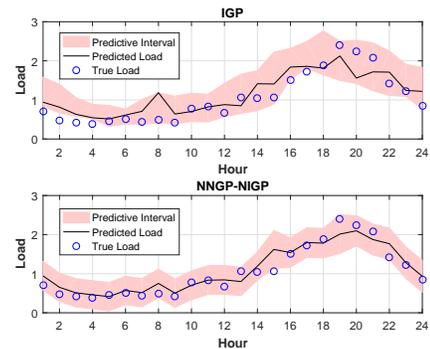}}
  \caption{The predictive intervals obtained by IGP and NNGP-NIGP in all test cases.}
  \label{fig:interval_com}
\end{figure}

The CPs obtained by NNGP-NIGP and IGP in all test cases are summarized in Fig. \ref{fig:box_cov_8bus}. We see that NNGP-NIGP has a higher coverage probability than IGP. To closely examine the ability of NNGP-NIGP in appropriately propagating the impact of input errors into the predictive uncertainty associated with the target customer's load forecasts, we show in Fig. \ref{fig:interval_com} the predictive intervals obtained by NNGP-NIGP and IGP for one arbitrarily chosen customer on a typical prediction day in all test cases. Notice that in Fig. \ref{fig:interval_com}, the predictive interval of NNGP-NIGP can cover more observations than IGP due to suitable input error propagation. Moreover, the predictive interval given by NNGP-NIGP is much narrower than that given by IGP. In fact, the predictive interval of IGP is found sometimes wider than actually needed, which may result in unnecessary waste of resources if used for planning and operation purposes. 

\subsection{Applying GSA to Improve Computational Efficiency }
We demonstrate that the proposed GSA method can effectively identify those important input features for load forecasting and improve the computational efficiency of NNGP-NIGP. 
Table \ref{table:alp} shows the estimated total Sobol' indices obtained  in the second stage of NNGP-NIGP  for load prediction of customer 1  on three different days in the 8-bus test case. Since the estimated Sobol' index value corresponding to the input $\theta_{1,2}$ is significantly higher than those for the other inputs,  $\theta_{1,2}$ is identified as the most important input for load prediction of customer 1. This conclusion is consistent with that of IGP given in \cite{Xie}, where a heuristic feature selection method relying on the automatic relevance determination kernel is used. As presenting the GSA results for the other test cases (especially for the 123-bus test case) would require considerable space, we mention without showing details that the GSA results are highly consistent with those obtained by IGP in other test cases as well. Table \ref{table:comp} shows the average computational times consumed by NNGP-NIGP with and without GSA in all test cases. We see that the computational time of NNGP-NIGP is significantly reduced when only those important input features are used for load forecasting.

\begin{table}[ht]
 \caption{Estimated total Sobol' indices $\widehat{S}^2_{T_j}$'s for the inputs $\theta_{1,2}, \theta_{1,3}, \ldots, \theta_{1,8}$ in the second stage of NNGP-NIGP. Predictions are performed for customer 1  on 3 days in the IEEE 8-bus test case.}
\label{table:alp}
\centering
	\begin{tabular}{ | c | c | c | c |c|}
	\hline
	    $\widehat{S}^2_{T_j}$ & $\theta$ & Day 1 &  Day 2 & Day 3 \\ \hline
		$\widehat{S}^2_{T_2}$ & $\theta_{1,2}$ & $0.834$ & $0.800$ & $0.897$\\ \hline
	    $\widehat{S}^2_{T_3}$ & $\theta_{1,3}$ & $0.015$ & $0.021$ & $0.002$ \\ \hline
	    $\widehat{S}^2_{T_4}$ & $\theta_{1,4}$ & $2.568\times10^{-10}$ & $0.084$ &
        $ 1.415\times10^{-9}$  \\ \hline
	    $\widehat{S}^2_{T_5}$ & $\theta_{1,5}$ & $3.435\times10^{-11}$ & $1.348\times10^{-9}$ & $1.558\times10^{-10}$   \\ \hline
	    $\widehat{S}^2_{T_6}$ & $\theta_{1,6}$ & $2.316\times10^{-16}$ & $9.519\times10^{-4}$& $9.984\times10^{-5}$ \\ \hline
	    $\widehat{S}^2_{T_7}$ & $\theta_{1,7}$ & $7.247\times10^{-11}$ & $1.460\times10^{-4}$ & 0.004 \\ \hline
	    $\widehat{S}^2_{T_8}$ & $\theta_{1,8}$ & $5.692\times10^{-11}$ & $2.690\times10^{-6}$ & $3.468\times10^{-10}$ \\ \hline
	\end{tabular}
\end{table}

\vskip-2ex
\begin{table}[ht]
 \caption{The average computational times (in seconds) of NNGP-NIGP with and without GSA for one day's prediction in all test cases.}
  \label{table:comp}
\centering
	\begin{tabular}{ | c | c | c|}
	\hline
	    Scale & NNGP-NIGP without GSA &  NNGP-NIGP with GSA  \\ \hline
		8-bus & $53.694$s & $2.469$s $\sim 4.228$s\\ \hline
		14-bus & $75.387$s & $2.247$s $\sim 12.375$s\\ \hline
		24-bus & $136.736$s & $3.104$s $\sim 43.583$s\\ \hline
	    123-bus & $241.381$s & $3.617$s $\sim 176.736$s \\ \hline
	\end{tabular}
\end{table}

\section{Conclusion}\label{sec:conclusion}
This work represents one of the first efforts to thoroughly address the impact of input errors on both point and interval predictions for the target customer's load. We identify and verify the opportunity of improving load forecasting performance by incorporating suitable input modeling and uncertainty quantification in a two-stage approach. The first stage delivers point and interval estimates for future input feature values, which are to be used in the second-stage model. The second stage propagates the impact of input errors into the ultimate point and interval predictions for the target customer's load. We select a representative two-stage load forecasting model, IGP, as the main benchmark model, which also performs input prediction, nevertheless via a much less sophisticated approach. The proposed two-stage approach is found to outperform IGP and a representative one-stage model, SARIMA, with respect to both point and interval predictions in various test cases. In addition, a model-free functional GSA method is considered to reduce the input-space dimensionality for an enhanced computational efficiency. The numerical results show that the GSA method can significantly improve the computational efficiency of the proposed two-stage approach.

\bibliographystyle{IEEEtran}
\bibliography{references}%

\newpage 
\begin{appendices} 
\section{Derivation of the equivalent GP for NNGP}\label{sec:app_NNGP}

In Section \ref{sec:nngp}, we have derived the equivalence between a one-layer neural network with infinite width and a GP. The covariance function of the corresponding GP is given in (\ref{eqn:cov_first}). In this section, we proceed to generalize the derivation to an NN with more than one layer (i.e., a DNN) by induction, and the key lies in deriving the covariance kernel of the equivalent GP that corresponds to a DNN. This step is particularly important for load forecasting, as it ensures that a DNN can be regarded as a GP, which can be conveniently used for obtaining point and interval estimates for future input values in a standard manner \cite{Ras2006GP}. Consider the $l$th layer $(l\geq 2)$ of the DNN, we have
\begin{eqnarray}
z_i^l(\mathbf{x})=b_i^l+\sum\limits_{j=1}^{N_l}W_{ij}^la_j^l(\mathbf{x}), \ \mbox{with} \ a_j^l(\mathbf{x})=\phi(z_j^{l-1}(\mathbf{x})).\nonumber
\end{eqnarray}
It follows that $z_i^l\sim \mathcal{N}(0,K^l)$, where $K^l$ is the covariance matrix in the $l$th layer, whose entry is given by the covariance between the $z_i^l$ values corresponding to a pair of inputs. For any two inputs $\mathbf{x}$ and $\mathbf{x}'$, we have  
\begin{eqnarray}\label{eqn:lthlayer}
K^l(\mathbf{x},\mathbf{x}')&=&\mathbb{E}[z_i^l(\mathbf{x})z_i^l(\mathbf{x}')]\nonumber \\ &=&\sigma^2_b+\sigma^2_w\mathbb{E}[\phi(z_i^{l-1}(\mathbf{x}))\phi(z_i^{l-1}(\mathbf{x}'))].
\end{eqnarray}
Notice that the joint distribution of $z_i^{l-1}(\mathbf{x})$ and $z_i^{l-1}(\mathbf{x}')$ is bivariate normal with zero mean and a $2\times 2$ covariance matrix whose distinct entries are given by $K^{l-1}(\mathbf{x},\mathbf{x}')$, $K^{l-1}(\mathbf{x},\mathbf{x})$, and $K^{l-1}(\mathbf{x}',\mathbf{x}')$. To show the recursive relationship between $K^l$ and $K^{l-1}$, we rewrite (\ref{eqn:lthlayer}) as 
\begin{eqnarray}\label{eqn:recursion}
K^l(\mathbf{x},\mathbf{x}')&= & \sigma^2_b+\sigma^2_wF_{\phi}(K^{l-1}(\mathbf{x},\mathbf{x}'),K^{l-1}(\mathbf{x},\mathbf{x}), \nonumber\\
&& K^{l-1}(\mathbf{x}',\mathbf{x}')),
\end{eqnarray}
where $F_{\phi}$ is a deterministic function that depends only on the nonlinearity function $\phi$. 
For the base case $K^0$, suppose that $W_{ij}^0\sim \mathcal{N}(0,\sigma_w^2/d_{in})$ and $b_j^0\sim \mathcal{N}(0,\sigma_b^2)$, we have 
\begin{eqnarray}\label{eqn:base}
K^0(\mathbf{x},\mathbf{x}')=\mathbb{E}[z_j^0(\mathbf{x})z_j^0(\mathbf{x}')]=\sigma_b^2+\sigma^2_w\left(\frac{\mathbf{x}\cdot \mathbf{x}'^\top}{d_{in}}\right).
\end{eqnarray}
The covariance matrix for the $l$th layer can thus be obtained recursively via (\ref{eqn:recursion}) and (\ref{eqn:base}). The resulting covariance matrix $K^L$ for the $L$th layer is the ultimate covariance kernel adopted by the NNGP model.  

To efficiently calculate the covariance kernel corresponding to the GP model equivalent to the DNN, an algorithm which utilizes the recursive relationship between the covariance matrix in adjacent layers (see (\ref{eqn:recursion})) is available. 

Constructing the covariance matrix $K^L$ for the equivalent GP involves computing the Gaussian integral in (\ref{eqn:lthlayer}) for all pairs
of training-training and training-test points, recursively for all layers. For some forms of nonlinearity functions used by the NNGP model, the integration can be performed analytically; but for most forms of nonlinearity functions, numerical integrations are required. Computing integrals independently for each pair of data points and each layer can be computationally expensive, and a more efficient algorithm to recursively calculate the covariance kernel $K^l$ in each layer $l$ is available in \cite{DNN}. We briefly summarize the algorithm as follows:

1. Generate: pre-activations $u = [−u_{max}, \ldots , u_{max}]$ consisting of $n_g$ elements evenly
spaced between $−u_{max}$ and $u_{max}$; variances $s = [0,s_{max}]$ with $n_v$ evenly spaced
elements, where $s_{max} < u^2_{max}$; and correlations $c = (-1, \ldots, 1)$ with $n_c$ linearly spaced
elements.

2. Populate a matrix $F$ containing a lookup table for the function $F_\phi$ given in (\ref{eqn:recursion}). This
involves numerically approximating a Gaussian integral, in terms of the marginal variances
$s$ and correlations $c$. 
The entries in the matrix $F$ are computed as

\begin{eqnarray}
F_{ij}=\frac{\sum\limits_{ab}\phi(u_a)\phi(u_b)\exp\left(-\frac{1}{2}U_{ab}^\top C^{-1} U_{ab} \right)}{\sum\limits_{ab}\exp\left(-\frac{1}{2}U_{ab}^\top C^{-1} U_{ab} \right)}, \nonumber
\end{eqnarray}
where $U_{ab}=[u_a,u_b]^\top$ with $u_a \in u$ and $u_b \in u$, and 
\begin{eqnarray}
C = \begin{bmatrix}
s_i & s_ic_j\\
s_ic_j & s_i   \nonumber
\end{bmatrix}.
\end{eqnarray}

3. For every pair of data points $\mathbf{x}$ and $\mathbf{x}'$ in layer $l$, compute $K^l(\mathbf{x}, \mathbf{x}')$ using (\ref{eqn:recursion}). Approximate the function $F_{\phi}(K^{l-1}(\mathbf{x},\mathbf{x}'),K^{l-1}(\mathbf{x},\mathbf{x}),K^{l-1}(\mathbf{x}',\mathbf{x}'))$ by bilinear interpolation into the matrix $F$ from Step 2, where we interpolate into $s_i$ using the value
of $K^{l-1}(\mathbf{x}, \mathbf{x})$, and interpolate into $c_j$ using
$K^{l-1}(\mathbf{x}, \mathbf{x}')/K^{l-1}(\mathbf{x}, \mathbf{x})$. Remember that
$K^{l-1}(\mathbf{x}, \mathbf{x}) = K^{l-1}(\mathbf{x}', \mathbf{x}')$, due to data pre-processing to guarantee constant norm.

4. Repeat the previous step recursively for all layers.

Following the algorithm above, the covariance of any two arbitrary inputs in the $L$th layer can be calculated efficiently. The resulting covariance matrix in the $L$th layer is then used for prediction for the corresponding GP. We refer the interested reader to \cite{DNN} for more details on this algorithm. A Python package is available for practical implementation of NNGP \cite{NNGPcode}.

\end{appendices}
\end{document}

%% file: math_macro.tex
\usepackage{amsmath,amssymb,amsfonts,graphicx,nicefrac,mathtools,bbm}
\usepackage{amsthm}

\newcommand{\BIT}{\begin{itemize}}
\newcommand{\EIT}{\end{itemize}}
\newcommand{\BNUM}{\begin{enumerate}}
\newcommand{\ENUM}{\end{enumerate}}
 
\def\mrm#1{\mathrm{#1}}

\providecommand{\diag}{\mathop\mathrm{diag}}
\providecommand{\tr}{\mathop\mathrm{tr}}

\def\Cov{\mrm{Cov}} 

